\documentclass[aps,prl,preprint,unsortedaddress,showkeys,nofootinbib]{revtex4-1}

\usepackage{amssymb,amsmath,amsfonts}

\usepackage{graphics}
\usepackage{graphicx}
\usepackage{epstopdf}
\usepackage[pdftex]{hyperref}
\usepackage{longtable}

\usepackage{bm}
\usepackage{subfigure}

\usepackage{color}
\usepackage{setspace}
\usepackage{multirow}

\usepackage[ruled,vlined,linesnumbered]{algorithm2e}

\usepackage{color, colortbl}
\definecolor{Yellow}{rgb}{1,1,0}
\definecolor{Grey}{rgb}{.87,.87,.87}
\definecolor{Purple}{rgb}{.8,.0,1.0}
\definecolor{Crimson}{rgb}{.86,.08,.23}

\begin{document}

\title{Optimality Of Community Structure In Complex Networks }

\author{Stanislav Sobolevsky
\footnote{To whom correspondence should be
addressed: sobolevsky@nyu.edu}}
\affiliation{
Center For Urban Science+Progress, New York University, Brooklyn, NY, USA\\
SENSEable City Laboratory,
Massachusetts Institute of Technology, Cambridge, MA, USA\\
Institute Of Design And Urban Studies of The National Research University ITMO, Saint-Petersburg, Russia
}
\author{Alexander Belyi}
\affiliation{Faculty of Applied Mathematics and Computer Science, Belarusian State University, Minsk, Belarus}
\affiliation{SENSEable City Laboratory, SMART Centre, Singapore}
\affiliation{SENSEable City Laboratory, Massachusetts Institute of Technology, Cambridge, MA, USA}
\author{Carlo Ratti}
\affiliation{SENSEable City Laboratory,
Massachusetts Institute of Technology, Cambridge, MA, USA}



\date{\today}

\begin{abstract}
\begin{it}
Community detection is one of the pivotal tools for discovering the structure of complex networks. Majority of community detection methods rely on optimization of certain quality functions characterizing the proposed community structure. Perhaps, the most commonly used of those quality functions is modularity. Many heuristics are claimed to be efficient in modularity maximization, which is usually justified in relative terms through comparison of their outcomes with those provided by other known algorithms. However as all the approaches are heuristics, while the complete brute-force is not feasible, there is no known way to understand if the obtained partitioning is really the optimal one. In this article we address the modularity maximization problem from the other side ---  finding an upper-bound estimate for the possible modularity values within a given network, allowing to better evaluate suggested community structures. Moreover, in some cases when then upper bound estimate meets the actually obtained modularity score, it provides a proof that the suggested community structure is indeed the optimal one. We propose an efficient algorithm for building such an upper-bound estimate and illustrate its usage on the examples of well-known classical and synthetic networks, being able to prove the optimality of the existing partitioning for some of the networks including well-known Zachary's Karate Club.
\end{it}
\end{abstract}

\keywords{Complex networks | Community detection | Network science}

\maketitle

\section*{Introduction}
A number of features of the increasingly interconnected world can be now described by means of complex networks - examples include but are not limited to physical and digital infrastructure, biological interactions, as well as human mobility and communications. The complex networks play crucial role in various fields such as physics, biology, economics, social sciences, urban planning. This gives paramount importance to the methods and approaches allowing to understand the underlying structure of the complex networks, in particular their community structure. 
The community detection saw a wide range of applications in social science \cite{plantie2013survey}, biology\cite{Guimera2005FunctionalCartography}, economics \cite{PiccardiWorldTradeWeb}. Some more specific applied examples include studies of human mobility and interactions with applications to regional delineation \cite{Ratti2010GB, blondel2010regions, Sobolevsky2013delineating, amini2014impact, hawelka2014geo, kang2013exploring, sobolevsky2014money, belyi2017global, grauwin2017identifying}.

Over the last 15 years a big number of approaches and algorithms for community detection in complex networks has been suggested. Some of them are just the straightforward heuristics such as hierarchical clustering\,\cite{Hastie2001ElementsOfStatisticalLearning} or the Girvan-Newman\,\cite{GN} algorithm, however the vast majority rely on optimization techniques based on the maximization of various objective functions. The first and the most well-known one is modularity\,\cite{newman2004,newman2006} assessing the relative strength of edges and quantifying the cumulative strength of the intra-community links. A large number of modularity optimization strategies have been suggested \cite{NewmanPRE2004, CNM2004VeryLargeNetworks, newman2004, newman2006, Sun2009, leuven, simulatedAnnealing,Good2010PerformanceOfModularity, Duch2005CElegans, LeeCSA, combo}. A good historical overview is presented in \cite{fortunato2010}.

Modularity is known to have certain shortcomings including a resolution limit\cite{Fortunato02012007ResolutionLimit, Good2010PerformanceOfModularity} and certain alternative objective functions deserve to be mentioned: Infomap description code length \cite{Rosvall01052007InformationTheoretic, Infomap}, block model likelihood \cite{Newman2011Stochastic,Newman2011Efficient,Bickel2009Nonparametric, Decelle2011BlockModel, Decelle2011BlockModelAsymptotics, Yan2012ModelSelection}, and surprise \cite{Aldecoa2011Deciphering}.

However despite its limitations modularity remains the most commonly used approach so far. Recently the authors proposed a novel universal optimization technique for community detection "Combo" \cite{combo} capable of maximizing modularity, description code length and pretty much any other metric based on the link scoring and assessing the cumulative score of the intra-community links (we will further refer to such metrics as link-scoring-based). For modularity optimization in most cases Combo outperforms other state-of-the-art algorithms in terms of the quality (modularity score) of the resulting partitioning which could be achieved within a reasonable time for the network of up to tens of thousands of nodes.

But whichever optimization technique is used, it is always a heuristic as the problem of finding exact modularity maximum is known to be NP-hard \cite{brandes2006maximizing} - the total number of partitioning to consider scales at the order of magnitude of $m^n$ where $m$ is the number of communities and $n$ is the size of the network. A heuristic is never capable of proving that the provided solution is indeed the optimal one and usually can not even tell how close to the optimal the solution is, with the rare exception of algorithms with some guaranteed performance estimates \cite{dinh2015network}. This way the partitioning quality is usually assessed in the relative terms by comparing the algorithm's outcome with the results of the other approaches. 

In the present paper we consider the problem from a different angle proposing an algorithm for building upper-bound estimate for the possible modularity score or any other link-scoring-based partitioning quality function. Sometimes, when such an estimate meets the value achieved by the optimization algorithm, this can provide a proof that the obtained solution is indeed the optimal one. We will provide examples of cases when such a proof becomes possible.

\section{The modularity function and its trivial upper bound}

If weights of network edges between each pair of nodes A,B are denoted as $e(A,B)$ then modularity~\cite{newman2004,newman2006} of the partition $c(A)$ (a mapping assigning community number $c$ to each node $A$) can be defined as 
\begin{equation}
Q=\sum_{A,B, c(A)=c(B)}q(A,B),
\label{modularity}
\end{equation}
where the quantities $q(A,B)$ for each edge $AB$ (call $q$ a modularity score for an edge) are defined as
$$
q(A,B)=\frac{e(A,B)}{T}-\frac{w^{out}(A)w^{in}(B)}{T^2},
$$
where $w^{out}(A)=\sum_C e(A,C)$, $w^{in}(B)=\sum_C e(C,B)$, $T=\sum_A w^{out}(A)=\sum_B w^{in}(B)=\sum_{A,B}e(A,B)$.
If the network is undirected then the edge modularity scores $q$ are symmetrical: $q(A,B)=q(B,A)$. However even for the directed case, the modularity scores could be effectively symmetrized assigning $q(A,B):=(q(A,B)+q(B,A))/2$ without any impact on the total score $Q$.

Consider a symmetrical square matrix of the edge modularity scores $q$. Worth mentioning that all the considerations below do not depend on the specific way the scores $q$ are defined, so are equally valid for any other objective function implemented through a cumulative edge scoring like (\ref{modularity}).

Modularity $Q$ of the complex network is known to be a normalized measure, i.e. $-1\leq Q\leq 1$. However can one ever reach the maximal value of $Q=1$? In most cases not and a trivial estimate of 
\begin{equation}
Q\leq Q^{max}=\sum_{A,B, q(A,B)>0}q(A,B)+\sum_{A, q(A,A)<0}q(A,A)
\label{qmax1}
\end{equation}
often leads to $Q^{max}<1$. But even this more precise upper bound (\ref{qmax1}) is often far from the actual values that could be ever achieved. Further we provide an algorithm for building a better estimate.

\section{The method}
Idea of the algorithm is based on the following consideration. Assume that the network contains three nodes $A,B,C$ such that the modularity scores $q(A,B)>0$ and $q(A,C)>0$, while $q(B,C)<0$. Then if we were to have both positive scores $q(A,B)$ and $q(A,C)$ as a part of the total modularity \ref{modularity}, i.e. have both links $AB$ and $AC$ as intra-community ones with respect to the considered community structure, all three nodes $A,B,C$ should belong to the same community. But this will come at a price of also having a link with a negative score $q(B,C)$ as intra-community one and would cause the corresponding negative score to be also included in \ref{modularity}. Otherwise at least one of the positive scores $q(A,B)$ or $q(A,C)$ would not be included. This way a triangle $ABC$ produces an additional "penalty" of $p(A,B,C)=2\min\{q(A,B),q(B,C),|q(A,C)|\}$ to the total modularity score estimate (the penalty is doubled as each edge score $q(A,B)$ is present in the network twice as $q(B,A)=q(A,B)$).

More generally, if a chain $A_1,A_2,...,A_k$ is such that $q(A_j,A_{j+1})>0$ for all $j=1,2,...,k-1$, while $q(A_1,A_k)<0$ then such a chain produces a penalty of 
\begin{equation}
p(A_1,A_2,...,A_k)=\min\{q(A_1,A_2),q(A_2,A_3),...,q(A_{k-1},A_k),|q(A_1,A_k)|\}.
\label{singlePenalty}
\end{equation} 
Call those triangles or chains {\it penalized}.

Based on that, if one finds a set $\Omega$ of non-overlapping (i.e. not having common edges) penalized undirected triangles or longer chains, then
\begin{equation}
Q\leq Q^{max}-\sum\limits_{\omega\in\Omega}p(\omega).
\label{qmax2}
\end{equation}

However, there is also a way to deal with the set of overlapping triangles or chains having common edges. Before we introduce the corresponding approach let us consider a more general framework which will make the idea more intuitive and will also enable further generalization of the algorithm.  

\section{Penalized subnetworks}

Start with several definitions. Recall that each edge AB of the original network is characterized by its modularity score $q(A,B)$ which can be positive, negative or zero.

\textit{Definition 1. Subnetwork (SN)}~-- a subset of nodes of the original network, where each edge A,B is characterized by a score $q^*(A,B)$ such that $|q^*(A,B)|\leq |q(A,B)|$ and $q^*(A,B)q(A,B)\geq 0$.

\textit{Definition 2. Subnetwork's penalty}~-- a difference between the sum of all edge scores $q^*$ of the subnetwork and the proven upper bound for its possible partitioning modularity estimate. Subnetwork with a positive penatly is called a penalized subnetwork.

\textit{Definition 3. Resolved subnetwork (RSN)}~-- a subnetwork, best known candidate partition of which matches the proved upper bound for its possible partitioning modularity estimate (making it a proven optimal partitioning).

\textit{Definition 4. Reduced subnetwork} of a given penalized subnetwork S ~-- a subnetwork of S having the same penalty as S.

\textit{Definition 5. Permissible linear combination} $LS=\sum_j \lambda_j S_j$ of the subnetworks $S_j$ with non-negative coefficients $\lambda_j$~-- is a subnetwork of the original network which edges AB characterized by the linear combinations $q^*(A,B)=\sum_j \lambda_j q^*_j(A,B)$ of the corresponding scores $q^*_j(A,B)$ from the subnetworks $S_j$ (in case AB is not present in some of the subnetworks consider corresponding $q^*_j(A,B)=0$), provided that $|q^*(A,B)|\leq |q(A,B)|$ and $q^*(A,B)q(A,B)\geq 0$.

Then it is easy to see that the following proposition holds:

{\bf Summation lemma}: Consider a set of subnetworks $\{S_1, S_2,..., S_m\}$ with proven penalties $p_1, p_2, ..., p_m$. Then then penalty for a permissible linear combination $LS=\sum_j \lambda_j S_j$ for non-negative $\lambda_j$ is at least $\sum_j \lambda_j p_j$.

Indeed, denote the proven upper bound estimate for the partitioning modularity for each subnetwork $S_j$ by $U_j$. Then for any given partitioning $c$ modularity $Q(c)$ of the $LS$ can be expressed as $Q=\sum_j \lambda_j Q_j$ and then $Q=\sum_j \lambda_j Q_j\leq \sum_j \lambda_j U_j=\sum_j \lambda_j (Q^{max}_j-p_j)=Q^{max}-\sum_j \lambda_j p_j$, where $Q^{max}$ as before denotes the estimate (\ref{qmax1}) for a corresponding subnetwork.

\textit{Definition 6. Resolved network/subnetwork} is a network or its subnetwork for which one was able to find a proven upper bound modularity estimate and an actual community structure with matching modularity scores.  

Resolving a network means constructing the proof of optimality for the known partitioning. Summation lemma provides a framework for resolving the networks through combining proven penalties of the smaller subnetworks.

\section{Overlapping triangles and chains}

Now return to an approach of handling a set of overlapping triangles or chains having common edges using summation rule.
Clearly penalized triangles or chains are examples of penalized subnetworks. If $p$ is the penalty then one can construct a reduced subnetwork of the same penalty with the same edges scored as $p$ and $-p$ depending on the sign of the original edge characteristics. 

Then considering any set $\Omega=\{\omega_j\}$ of overlapping penalized triangles or chains $\omega_j$, one can replace it by a set of the corresponding reduced penalized triangles or chains $\overline{\omega_j}$. If a linear combination of those is a permissible linear combination (i.e. if a linear combination of penalties of all the triangles/chains each edge participates in does not exceed the absolute value of the modularity score of this edge) then summation rule provides a proven cumulative penalty for the entire network as a linear combination of the penalties of all the triangles and chains. 

In particular if $\sum_j\overline{\omega_j}$ is a permissible linear combination (with $\lambda_j=1$) then
\begin{equation}
Q\leq Q^{max}-\sum\limits_j p(\omega_j).
\label{qmax3}
\end{equation}
This way we replace overlapping penalized triangles and chains with non-overlapping (in terms of the edge scores) reduced penalized triangles and chains.

\section{The algorithm}
An upper bound estimate (\ref{qmax3}) largely depends on the set $\Omega$ of triangles or chains.
Introduce a greedy heuristic building a sequence of triangles and chains $\Omega=\{\omega_1,\omega_2,...,\omega_m\}$ aiming to produce the best possible upper bound estimate (\ref{qmax3}):
\begin{enumerate}
    \item Start from $k=3$ and empty $\Omega$.
    \item Look for all the penalized chains of length $k$: search through the negatively scored edges $AB$, considering all the paths between $A$ and $B$ which consist of all positively scored edges, evaluating penalty for every such chain.
    \item If there exists at least one such chain, add to $\Omega$ the one having highest associated penalty value $p(\omega)$; reduce the edge scores of the original network by subtracting the corresponding edge scores $p(\omega)$ or $-p(\omega)$ of the reduced chain $\overline{\omega}$; repeat from the step 2.
    \item Otherwise check if there still exist any penalized chain (evaluate connected components in the graph with edges having positive residual score and see if at least one of them contains an internal edge with the negative residual score). If such a chain exists increment $k$ and repeat from step 2. 
    \item Otherwise stop.
\end{enumerate}

The algorithm is streightforward and relatively fast, however as pretty much any greedy heuristic sometimes converges to the non-optimal solutions. Alternative for the step 3 is to select a random penalized chain at each step. Also one can select the best chain with a certain probability and a random chain otherwise. In those cases each series of iterations 2-3 for each $k$ could be performed multiple times in order to select the best intermediate result before incrementing $k$.

\section{The results}
First we provide in Table~\ref{tab:Results1} several examples of the networks for which we are able to prove that the obtained best modulairity partitioning is indeed the optimal one. Here and further we use Combo algorithm for modularity optimization.

\begin{longtable*}[c]{| r | c | c | c | c |}
\hline
Network & Size & Acheived modularity & Upper-bound modularity & Ratio, \%\\
\hline
\endfirsthead
\hline
\endhead
\hline
\endfoot
Ants 1 \cite{cole1981dominance}& 16 & 0.0972 & 0.0972 & 100.0\\
Ants 2 \cite{cole1981dominance}& 13 & 0.1666 & 0.1666 & 100.0\\
Dolphin groups \cite{connor1992dolphin} & 13 & 0.5900 & 0.5900 & 100.0\\
Pony \cite{clutton1976ranks} & 17 & 0.1375 & 0.1375 & 100.0\\
GAMAPOS \cite{UCINETIV}& 16 & 0.525565 & 0.525565 & 100.0\\
PADGM \cite{UCINETIV}& 16 & 0.364444 & 0.364444 & 100.0 \\
KNOKM \cite{UCINETIV}& 10 & 0.14876 & 0.14876 & 100.0 \\
KNOKI \cite{UCINETIV}& 10 & 0.0816327 & 0.0816327 & 100.0\\
Synthetic network & 50 & 0.6399 & 0.6399 & 100.0\\
\hline
\caption{Evaluation on the real-world and synthetic networks: examples of the networks with exact or nearly exact match between upper-bound estimate and the achieved result. The synthetic network was generated using LFR approach \cite{LFR}, while real-world network taken from \cite{Moreno} and \cite{UCINETIV}\label{tab:Results1}}
\end{longtable*}

Second we perform extensive evaluations of the algorithm on a number of networks of size 25-75 created using artificial network generator of Lancichinetti- Fortunato- Radicchi (LFR) benchmark \cite{LFR} with
average degree equal $4$,
maximum degree $= 16$,
mixing parameter $\mu=0.1-0.2$ for the topology equal to mixing parameter for the weights,
exponent for the weight distribution $= 1$,
minus exponent for the degree sequence $= 2$,
minus exponent for the community size distribution $= 1$,
minimum for the community sizes $= 5$,
maximum for the community sizes equal to number of nodes divided by $4$.

Results are reported in the Table~\ref{tab:Results2} 
\begin{longtable*}[c]{| r | c | c | c | c |}
\hline
Network size & Avg. ratio of achieved vs upper bound modularity, \% & Exact matches, \%\\
\hline
\endfirsthead
\hline
\endhead
\hline
\endfoot
25-34 & 100.0 & 100.0\\
35-44 & 99.96 & 80.0\\
45-54 & 99.98 & 60.0\\
55-64 & 99.91 & 50.0\\
65-75 & 99.82 & 18.18\\
All & 99.93 & 60.78\\
\hline
\caption{Evaluation on the synthetic LFR networks.\label{tab:Results2}}
\end{longtable*}

Further in the Table~\ref{tab:Results3} we present results for the other real-world networks where exact match between the upper bound estimate and the best known modularity score has not been achieved. In most cases the gap between the two is below $5\%$, often (if half of those cases) within $0.1-1\%$.

\begin{longtable*}[c]{| r | c | c | c | c | c |}
\hline
\# & Network & Size & Achieved modularity & Upper-bound modularity & Ratio, \%\\
\hline
\endfirsthead
\hline
\endhead
\hline
\endfoot
1  & Zachary's karate club \cite{zachary1977ifm} & 34    &  0.419790 & 0.425789 & 98.59\\
2  & Dolphins \cite{Lusseau2003Dolphins} & 62    &  0.528519 & 0.548080 & 96.43\\
3  & Les Miserables \cite{Knuth1993GraphBase} & 77    &  0.566688 & 0.572035 & 99.07\\
4  & Political Books [\hyperref{http://www.orgnet.com}{}{}{orgnet.com}] & 105   &  0.527237 & 0.541007 & 97.45\\
5  & American College Football \cite{GN} & 115   &  0.605445 & 0.62767 & 96.46\\
6 & World colonial ties \cite{belyi2017global}& 238 & 0.7874 & 0.7875 & 99.99\\ 
7 & World language ties \cite{belyi2017global} & 238 & 0.3714 & 0.3717 & 99.92\\ 
8  & Neural net of C. Elegans \cite{White12111986} & 297   &  0.507642 & 0.541827 & 93.69\\
9 & Infectious diseases \cite{isella2011s} & 410 &  0.860384 & 0.86777 & 99.15\\
10 & Synthetic network & 250   &  0.798493 & 0.79922 & 99.91\\
11 & Synthetic network & 500   &  0.851837 & 0.85584 & 99.53\\ 
12 & Jazz musicians \cite{Gleiser2003Jazz} & 198   &  0.444787 & 0.46885 & 94.87\\
13 & WOLFN \cite{UCINETIV}& 20 & 0.0580522 & 0.0626775 & 92.62\\
14 & THURM \cite{UCINETIV} & 15 & 0.269513 & 0.275023 & 98.00 \\
15 & GAMANEG \cite{UCINETIV}& 16 & 0.274673 & 0.3044 & 90.23 \\
16 & PADGB \cite{UCINETIV}& 16 & 0.39875 & 0.415 & 96.08 \\
17 & KAPFMM \cite{UCINETIV}& 15 & 0.278393 & 0.295014 & 94.37 \\

\hline
\caption{Evaluation on the real-world networks: best and upper-bound modularity score for the extensive set of networks.\label{tab:Results3}}
\end{longtable*}

\section{Further improvements of the algorithm}

\begin{figure}[b]
\centering
\includegraphics[width=0.5\textwidth]{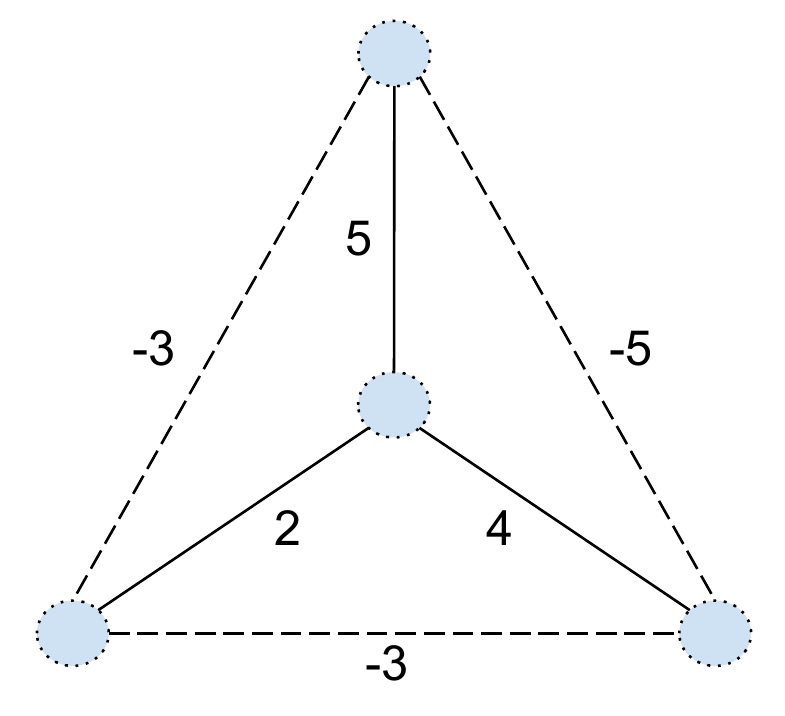}
\caption{\label{fig::Unresolvable}Example of the network configuration for which exact upper bound on modularity can not be found by algorithm considering triangles and chains. Numbers indicate scaled edges' modularity scores.}
\end{figure}

As we see the presented algorithm can build a plausible upper-bound estimate for the possible modularity score for many real-world and synthetic networks, often being able to resolve them (i.e. matching the modularity score achievable by the community detection algorithm (Combo), proving the optimality of such a partitioning). However, there are networks that fundamentally can not be resolved with the above approach - see Figure~\ref{fig::Unresolvable} for an example of a simple unresolvable network configuration. And despite the fact that the gap between the proven upper bound estimate and achievable modularity score might be as small as $0.1\%$ or even less this still does not allow to prove the optimality of the partitioning which is seen as the ultimate goal of the present paper.

In order to overcome the fundamental limitations of the approach based on penalized triangles and chains we propose a more general framework able to utilize arbitrary penalized subnetworks, including the one on the Figure~\ref{fig::Unresolvable}.  

{\bf Scheme of the framework}:\\
Phase 1. Identify a set of arbitrary subnetworks $S_1, S_2,..., S_m$ with known penalties $p_1, p_2, ..., p_m$.\\
Phase 2. Find a best possible acceptable linear combination $\sum_j \lambda_j S_j$, such that $\sum_j \lambda_j p_j\to \max$.

Instead of a binary (i.e. having $\lambda_j=0,1$) inclusion heuristic like we tried before, Phase 2 could be implemented as a linear optimization problem maximizing $\sum_j \lambda_j p_j\to {\rm max}$ subject to linear constrains $\lambda_j\geq 0$ and the corresponding constrains on each edge score, ensuring that the resulting linear combination of the subnetworks is permissible. 

The phase 1 requires us to decide which $S_j$ to include and how to build them. A simple although perhaps not time efficient generic solution could consider all possible subnetworks of a small size up to a certain size $M$. For $M$ small enough all such possible subnetworks could be analyzed within a reasonable time. 
Each subnetwork can be resolved using the following algorithm, which we call \textbf{\textit{partial brute-force}}:
\begin{enumerate}
    \item Start with $m:=0$ and with the upper bound modularity estimate $Q^*:=0$;
    \item Increment $m$ ($m:=m+1$) if $m<M$, otherwise - stop, marking the subnetwork as unresolved;
    \item Consider all combinations of excluding exactly $m$ edges and merging nodes connected by other positive edges into communities (those combinations where edges to merge along imply merging along some of those other edges which were intended to be excluded could be discarded as they have to be already considered as the ones with smaller $m$);
    \item For each of the combinations above assess the resulting modularity $Q_1$ and set $Q^*:=\max{(Q^*, Q_1)}$;
    \item If sum of all positive edges but the $m+1$ smallest ones is higher than $Q^*$, return to step 2, otherwise stop marking the subnetwork as resolved.
\end{enumerate}

Once the algorithm finalizes marking subnetwork as resolved, then $p=Q^{max}-Q^*$ is the proven penalty.

And then for each penalized subnetwork resolved with the above approach we try to reduce the weights of its edges as much as possible provided that the penalty is preserved. This is done by solving a linear programming problem minimizing the sum of the edge weights with respect to the following conditions:
\begin{enumerate}
    \item For each partition considered by the procedure above, sum of the absolute values of edge scores  contributing to penalty (positive edges between communities and negative edges within communities) can not be less than $p$;
    \item For each set of the $m+1$ edges with positive scores, sum of their scores can not be less than $p$.
\end{enumerate}

The above two procedures applied to all subnetworks up to a certain size $M$ provide a set of reduced resolved subnetworks, completing phase 1 of the framework. Then a standard linear programming solution like a simplex method could be used to implement phase 2.

Current implementation of this approach for $M=6$ for example allows us to improve the earlier obtained upper bound modularity estimates, resolving a famous Zachary's karate club network \cite{zachary1977ifm} as well as the networks 13-17 from the Table~\ref{tab:Results3}. This way we find exact match between the best modularity score (e.g. 0.41979 for Zachary's karate club) achieved by Combo and the upper-bound estimate we provide proving the optimality of the obtained partitioning.
We believe the approach might allow resolving much broader sets of the networks, however implementation requires further optimization so that it can efficiently handle networks of larger size within a reasonable time.

\section{Conclusions}
We proposed an efficient algorithm for building an upper-bound estimate for the possible partitioning modularity score of a network. The algorithm is applied to a set of well-known classical and synthetic networks, being able to prove the optimality of the existing partitioning for some of the networks including well-known Zachary's Karate Club. For the rest of the networks the constructed upper bound modularity estimates happen to be pretty close to the achieved scores of the known partitioning (the gap between the achieved score and the upper-bound estimate is often within $1\%$ of the score). The ways of further improving the approach are provided.


\begin{thebibliography}{49}
\expandafter\ifx\csname natexlab\endcsname\relax\def\natexlab#1{#1}\fi
\expandafter\ifx\csname bibnamefont\endcsname\relax
  \def\bibnamefont#1{#1}\fi
\expandafter\ifx\csname bibfnamefont\endcsname\relax
  \def\bibfnamefont#1{#1}\fi
\expandafter\ifx\csname citenamefont\endcsname\relax
  \def\citenamefont#1{#1}\fi
\expandafter\ifx\csname url\endcsname\relax
  \def\url#1{\texttt{#1}}\fi
\expandafter\ifx\csname urlprefix\endcsname\relax\def\urlprefix{URL }\fi
\providecommand{\bibinfo}[2]{#2}
\providecommand{\eprint}[2][]{\url{#2}}

\bibitem[{\citenamefont{Planti{\'e} and Crampes}(2013)}]{plantie2013survey}
\bibinfo{author}{\bibfnamefont{M.}~\bibnamefont{Planti{\'e}}} \bibnamefont{and}
  \bibinfo{author}{\bibfnamefont{M.}~\bibnamefont{Crampes}}, in
  \emph{\bibinfo{booktitle}{Social media retrieval}}
  (\bibinfo{publisher}{Springer}, \bibinfo{year}{2013}), pp.
  \bibinfo{pages}{65--85}.

\bibitem[{\citenamefont{Guimer\`a and
  Nunes~Amaral}(2005)}]{Guimera2005FunctionalCartography}
\bibinfo{author}{\bibfnamefont{R.}~\bibnamefont{Guimer\`a}} \bibnamefont{and}
  \bibinfo{author}{\bibfnamefont{L.~A.} \bibnamefont{Nunes~Amaral}},
  \bibinfo{journal}{Nature} \textbf{\bibinfo{volume}{433}},
  \bibinfo{pages}{895} (\bibinfo{year}{2005}), ISSN \bibinfo{issn}{0028-0836},
  \urlprefix\url{http://dx.doi.org/10.1038/nature03288}.

\bibitem[{\citenamefont{Piccardi and Tajoli}(2012)}]{PiccardiWorldTradeWeb}
\bibinfo{author}{\bibfnamefont{C.}~\bibnamefont{Piccardi}} \bibnamefont{and}
  \bibinfo{author}{\bibfnamefont{L.}~\bibnamefont{Tajoli}},
  \bibinfo{journal}{Phys. Rev. E} \textbf{\bibinfo{volume}{85}},
  \bibinfo{pages}{066119} (\bibinfo{year}{2012}),
  \urlprefix\url{http://link.aps.org/doi/10.1103/PhysRevE.85.066119}.

\bibitem[{\citenamefont{Ratti et~al.}(2010)\citenamefont{Ratti, Sobolevsky,
  Calabrese, Andris, Reades, Martino, Claxton, and Strogatz}}]{Ratti2010GB}
\bibinfo{author}{\bibfnamefont{C.}~\bibnamefont{Ratti}},
  \bibinfo{author}{\bibfnamefont{S.}~\bibnamefont{Sobolevsky}},
  \bibinfo{author}{\bibfnamefont{F.}~\bibnamefont{Calabrese}},
  \bibinfo{author}{\bibfnamefont{C.}~\bibnamefont{Andris}},
  \bibinfo{author}{\bibfnamefont{J.}~\bibnamefont{Reades}},
  \bibinfo{author}{\bibfnamefont{M.}~\bibnamefont{Martino}},
  \bibinfo{author}{\bibfnamefont{R.}~\bibnamefont{Claxton}}, \bibnamefont{and}
  \bibinfo{author}{\bibfnamefont{S.~H.} \bibnamefont{Strogatz}},
  \bibinfo{journal}{PLoS ONE} \textbf{\bibinfo{volume}{5}},
  \bibinfo{pages}{e14248} (\bibinfo{year}{2010}),
  \urlprefix\url{http://dx.doi.org/10.1371%2Fjournal.pone.0014248}.

\bibitem[{\citenamefont{Blondel et~al.}(2010)\citenamefont{Blondel, Krings, and
  Thomas}}]{blondel2010regions}
\bibinfo{author}{\bibfnamefont{V.}~\bibnamefont{Blondel}},
  \bibinfo{author}{\bibfnamefont{G.}~\bibnamefont{Krings}}, \bibnamefont{and}
  \bibinfo{author}{\bibfnamefont{I.}~\bibnamefont{Thomas}},
  \bibinfo{journal}{Brussels Studies. La revue scientifique {\'e}lectronique
  pour les recherches sur Bruxelles/Het elektronisch wetenschappelijk
  tijdschrift voor onderzoek over Brussel/The e-journal for academic research
  on Brussels}  (\bibinfo{year}{2010}).

\bibitem[{\citenamefont{Sobolevsky et~al.}(2013)\citenamefont{Sobolevsky,
  Szell, Campari, Couronn{\'e}, Smoreda, and
  Ratti}}]{Sobolevsky2013delineating}
\bibinfo{author}{\bibfnamefont{S.}~\bibnamefont{Sobolevsky}},
  \bibinfo{author}{\bibfnamefont{M.}~\bibnamefont{Szell}},
  \bibinfo{author}{\bibfnamefont{R.}~\bibnamefont{Campari}},
  \bibinfo{author}{\bibfnamefont{T.}~\bibnamefont{Couronn{\'e}}},
  \bibinfo{author}{\bibfnamefont{Z.}~\bibnamefont{Smoreda}}, \bibnamefont{and}
  \bibinfo{author}{\bibfnamefont{C.}~\bibnamefont{Ratti}},
  \bibinfo{journal}{PloS ONE} \textbf{\bibinfo{volume}{8}},
  \bibinfo{pages}{e81707} (\bibinfo{year}{2013}).

\bibitem[{\citenamefont{Amini et~al.}(2014)\citenamefont{Amini, Kung, Kang,
  Sobolevsky, and Ratti}}]{amini2014impact}
\bibinfo{author}{\bibfnamefont{A.}~\bibnamefont{Amini}},
  \bibinfo{author}{\bibfnamefont{K.}~\bibnamefont{Kung}},
  \bibinfo{author}{\bibfnamefont{C.}~\bibnamefont{Kang}},
  \bibinfo{author}{\bibfnamefont{S.}~\bibnamefont{Sobolevsky}},
  \bibnamefont{and} \bibinfo{author}{\bibfnamefont{C.}~\bibnamefont{Ratti}},
  \bibinfo{journal}{EPJ Data Science} \textbf{\bibinfo{volume}{3}},
  \bibinfo{pages}{6} (\bibinfo{year}{2014}).

\bibitem[{\citenamefont{Hawelka et~al.}(2014)\citenamefont{Hawelka, Sitko,
  Beinat, Sobolevsky, Kazakopoulos, and Ratti}}]{hawelka2014geo}
\bibinfo{author}{\bibfnamefont{B.}~\bibnamefont{Hawelka}},
  \bibinfo{author}{\bibfnamefont{I.}~\bibnamefont{Sitko}},
  \bibinfo{author}{\bibfnamefont{E.}~\bibnamefont{Beinat}},
  \bibinfo{author}{\bibfnamefont{S.}~\bibnamefont{Sobolevsky}},
  \bibinfo{author}{\bibfnamefont{P.}~\bibnamefont{Kazakopoulos}},
  \bibnamefont{and} \bibinfo{author}{\bibfnamefont{C.}~\bibnamefont{Ratti}},
  \bibinfo{journal}{Cartography and Geographic Information Science}
  \textbf{\bibinfo{volume}{41}}, \bibinfo{pages}{260} (\bibinfo{year}{2014}).

\bibitem[{\citenamefont{Kang et~al.}(2013)\citenamefont{Kang, Sobolevsky, Liu,
  and Ratti}}]{kang2013exploring}
\bibinfo{author}{\bibfnamefont{C.}~\bibnamefont{Kang}},
  \bibinfo{author}{\bibfnamefont{S.}~\bibnamefont{Sobolevsky}},
  \bibinfo{author}{\bibfnamefont{Y.}~\bibnamefont{Liu}}, \bibnamefont{and}
  \bibinfo{author}{\bibfnamefont{C.}~\bibnamefont{Ratti}}, in
  \emph{\bibinfo{booktitle}{Proceedings of the 2nd ACM SIGKDD International
  Workshop on Urban Computing}} (\bibinfo{organization}{ACM},
  \bibinfo{year}{2013}), p.~\bibinfo{pages}{1}.

\bibitem[{\citenamefont{Sobolevsky
  et~al.}(2014{\natexlab{a}})\citenamefont{Sobolevsky, Sitko, Des~Combes,
  Hawelka, Arias, and Ratti}}]{sobolevsky2014money}
\bibinfo{author}{\bibfnamefont{S.}~\bibnamefont{Sobolevsky}},
  \bibinfo{author}{\bibfnamefont{I.}~\bibnamefont{Sitko}},
  \bibinfo{author}{\bibfnamefont{R.~T.} \bibnamefont{Des~Combes}},
  \bibinfo{author}{\bibfnamefont{B.}~\bibnamefont{Hawelka}},
  \bibinfo{author}{\bibfnamefont{J.~M.} \bibnamefont{Arias}}, \bibnamefont{and}
  \bibinfo{author}{\bibfnamefont{C.}~\bibnamefont{Ratti}}, in
  \emph{\bibinfo{booktitle}{Big Data (BigData Congress), 2014 IEEE
  International Congress on}} (\bibinfo{organization}{IEEE},
  \bibinfo{year}{2014}{\natexlab{a}}), pp. \bibinfo{pages}{136--143}.

\bibitem[{\citenamefont{Belyi et~al.}(2017)\citenamefont{Belyi, Bojic,
  Sobolevsky, Sitko, Hawelka, Rudikova, Kurbatski, and
  Ratti}}]{belyi2017global}
\bibinfo{author}{\bibfnamefont{A.}~\bibnamefont{Belyi}},
  \bibinfo{author}{\bibfnamefont{I.}~\bibnamefont{Bojic}},
  \bibinfo{author}{\bibfnamefont{S.}~\bibnamefont{Sobolevsky}},
  \bibinfo{author}{\bibfnamefont{I.}~\bibnamefont{Sitko}},
  \bibinfo{author}{\bibfnamefont{B.}~\bibnamefont{Hawelka}},
  \bibinfo{author}{\bibfnamefont{L.}~\bibnamefont{Rudikova}},
  \bibinfo{author}{\bibfnamefont{A.}~\bibnamefont{Kurbatski}},
  \bibnamefont{and} \bibinfo{author}{\bibfnamefont{C.}~\bibnamefont{Ratti}},
  \bibinfo{journal}{International Journal of Geographical Information Science}
  \textbf{\bibinfo{volume}{31}}, \bibinfo{pages}{1381} (\bibinfo{year}{2017}).

\bibitem[{\citenamefont{Grauwin et~al.}(2017)\citenamefont{Grauwin, Szell,
  Sobolevsky, H{\"o}vel, Simini, Vanhoof, Smoreda, Barab{\'a}si, and
  Ratti}}]{grauwin2017identifying}
\bibinfo{author}{\bibfnamefont{S.}~\bibnamefont{Grauwin}},
  \bibinfo{author}{\bibfnamefont{M.}~\bibnamefont{Szell}},
  \bibinfo{author}{\bibfnamefont{S.}~\bibnamefont{Sobolevsky}},
  \bibinfo{author}{\bibfnamefont{P.}~\bibnamefont{H{\"o}vel}},
  \bibinfo{author}{\bibfnamefont{F.}~\bibnamefont{Simini}},
  \bibinfo{author}{\bibfnamefont{M.}~\bibnamefont{Vanhoof}},
  \bibinfo{author}{\bibfnamefont{Z.}~\bibnamefont{Smoreda}},
  \bibinfo{author}{\bibfnamefont{A.-L.} \bibnamefont{Barab{\'a}si}},
  \bibnamefont{and} \bibinfo{author}{\bibfnamefont{C.}~\bibnamefont{Ratti}},
  \bibinfo{journal}{Scientific Reports} \textbf{\bibinfo{volume}{7}}
  (\bibinfo{year}{2017}).

\bibitem[{\citenamefont{Hastie}(2001)}]{Hastie2001ElementsOfStatisticalLearning}
\bibinfo{author}{\bibfnamefont{T.}~\bibnamefont{Hastie}},
  \emph{\bibinfo{title}{The elements of statistical learning : data mining,
  inference, and prediction : with 200 full-color illustrations}}
  (\bibinfo{publisher}{Springer}, \bibinfo{address}{New York},
  \bibinfo{year}{2001}), ISBN \bibinfo{isbn}{0387952845}.

\bibitem[{\citenamefont{Girvan and Newman}(2002)}]{GN}
\bibinfo{author}{\bibfnamefont{M.}~\bibnamefont{Girvan}} \bibnamefont{and}
  \bibinfo{author}{\bibfnamefont{M.}~\bibnamefont{Newman}},
  \bibinfo{journal}{Proc. Natl. Acad. Sci. USA} \textbf{\bibinfo{volume}{99
  (12)}}, \bibinfo{pages}{7821} (\bibinfo{year}{2002}).

\bibitem[{\citenamefont{Newman and Girvan}(2004)}]{newman2004}
\bibinfo{author}{\bibfnamefont{M.}~\bibnamefont{Newman}} \bibnamefont{and}
  \bibinfo{author}{\bibfnamefont{M.}~\bibnamefont{Girvan}},
  \bibinfo{journal}{Phys. Rev. E} \textbf{\bibinfo{volume}{69 (2)}},
  \bibinfo{pages}{026113} (\bibinfo{year}{2004}).

\bibitem[{\citenamefont{Newman}(2006)}]{newman2006}
\bibinfo{author}{\bibfnamefont{M.}~\bibnamefont{Newman}},
  \bibinfo{journal}{Proceedings of the National Academy of Sciences}
  \textbf{\bibinfo{volume}{103}}, \bibinfo{pages}{8577} (\bibinfo{year}{2006}).

\bibitem[{\citenamefont{Newman}(2004)}]{NewmanPRE2004}
\bibinfo{author}{\bibfnamefont{M.~E.~J.} \bibnamefont{Newman}},
  \bibinfo{journal}{Phys. Rev. E} \textbf{\bibinfo{volume}{69}},
  \bibinfo{pages}{066133} (\bibinfo{year}{2004}),
  \urlprefix\url{http://link.aps.org/doi/10.1103/PhysRevE.69.066133}.

\bibitem[{\citenamefont{Clauset et~al.}(2004)\citenamefont{Clauset, Newman, and
  Moore}}]{CNM2004VeryLargeNetworks}
\bibinfo{author}{\bibfnamefont{A.}~\bibnamefont{Clauset}},
  \bibinfo{author}{\bibfnamefont{M.~E.~J.} \bibnamefont{Newman}},
  \bibnamefont{and} \bibinfo{author}{\bibfnamefont{C.}~\bibnamefont{Moore}},
  \bibinfo{journal}{Phys. Rev. E} \textbf{\bibinfo{volume}{70}},
  \bibinfo{pages}{066111} (\bibinfo{year}{2004}),
  \urlprefix\url{http://link.aps.org/doi/10.1103/PhysRevE.70.066111}.

\bibitem[{\citenamefont{Sun et~al.}(2009)\citenamefont{Sun, Danila, Josi{\'c},
  and Bassler}}]{Sun2009}
\bibinfo{author}{\bibfnamefont{Y.}~\bibnamefont{Sun}},
  \bibinfo{author}{\bibfnamefont{B.}~\bibnamefont{Danila}},
  \bibinfo{author}{\bibfnamefont{K.}~\bibnamefont{Josi{\'c}}},
  \bibnamefont{and} \bibinfo{author}{\bibfnamefont{K.~E.}
  \bibnamefont{Bassler}}, \bibinfo{journal}{EPL (Europhysics Letters)}
  \textbf{\bibinfo{volume}{86}}, \bibinfo{pages}{28004} (\bibinfo{year}{2009}),
  \urlprefix\url{http://stacks.iop.org/0295-5075/86/i=2/a=28004}.

\bibitem[{\citenamefont{Blondel et~al.}(2008)\citenamefont{Blondel, Guillaume,
  Lambiotte, and Lefebvre}}]{leuven}
\bibinfo{author}{\bibfnamefont{V.~D.} \bibnamefont{Blondel}},
  \bibinfo{author}{\bibfnamefont{J.-L.} \bibnamefont{Guillaume}},
  \bibinfo{author}{\bibfnamefont{R.}~\bibnamefont{Lambiotte}},
  \bibnamefont{and} \bibinfo{author}{\bibfnamefont{E.}~\bibnamefont{Lefebvre}},
  \bibinfo{journal}{Journal of Statistical Mechanics: Theory and Experiment}
  \textbf{\bibinfo{volume}{2008}}, \bibinfo{pages}{P10008}
  (\bibinfo{year}{2008}).

\bibitem[{\citenamefont{Guimera et~al.}(2004)\citenamefont{Guimera,
  Sales-Pardo, and Amaral}}]{simulatedAnnealing}
\bibinfo{author}{\bibfnamefont{R.}~\bibnamefont{Guimera}},
  \bibinfo{author}{\bibfnamefont{M.}~\bibnamefont{Sales-Pardo}},
  \bibnamefont{and} \bibinfo{author}{\bibfnamefont{L.~A.~N.}
  \bibnamefont{Amaral}}, \bibinfo{journal}{Physical Review E}
  \textbf{\bibinfo{volume}{70}}, \bibinfo{pages}{025101}
  (\bibinfo{year}{2004}).

\bibitem[{\citenamefont{Good et~al.}(2010)\citenamefont{Good, de~Montjoye, and
  Clauset}}]{Good2010PerformanceOfModularity}
\bibinfo{author}{\bibfnamefont{B.~H.} \bibnamefont{Good}},
  \bibinfo{author}{\bibfnamefont{Y.-A.} \bibnamefont{de~Montjoye}},
  \bibnamefont{and} \bibinfo{author}{\bibfnamefont{A.}~\bibnamefont{Clauset}},
  \bibinfo{journal}{Phys. Rev. E} \textbf{\bibinfo{volume}{81}},
  \bibinfo{pages}{046106} (\bibinfo{year}{2010}),
  \urlprefix\url{http://link.aps.org/doi/10.1103/PhysRevE.81.046106}.

\bibitem[{\citenamefont{Duch and Arenas}(2005)}]{Duch2005CElegans}
\bibinfo{author}{\bibfnamefont{J.}~\bibnamefont{Duch}} \bibnamefont{and}
  \bibinfo{author}{\bibfnamefont{A.}~\bibnamefont{Arenas}},
  \bibinfo{journal}{Phys. Rev. E} \textbf{\bibinfo{volume}{72}},
  \bibinfo{pages}{027104} (\bibinfo{year}{2005}),
  \urlprefix\url{http://link.aps.org/doi/10.1103/PhysRevE.72.027104}.

\bibitem[{\citenamefont{Lee et~al.}(2012)\citenamefont{Lee, Gross, and
  Lee}}]{LeeCSA}
\bibinfo{author}{\bibfnamefont{J.}~\bibnamefont{Lee}},
  \bibinfo{author}{\bibfnamefont{S.~P.} \bibnamefont{Gross}}, \bibnamefont{and}
  \bibinfo{author}{\bibfnamefont{J.}~\bibnamefont{Lee}},
  \bibinfo{journal}{Phys. Rev. E} \textbf{\bibinfo{volume}{85}},
  \bibinfo{pages}{056702} (\bibinfo{year}{2012}),
  \urlprefix\url{http://link.aps.org/doi/10.1103/PhysRevE.85.056702}.

\bibitem[{\citenamefont{Sobolevsky
  et~al.}(2014{\natexlab{b}})\citenamefont{Sobolevsky, Campari, Belyi, and
  Ratti}}]{combo}
\bibinfo{author}{\bibfnamefont{S.}~\bibnamefont{Sobolevsky}},
  \bibinfo{author}{\bibfnamefont{R.}~\bibnamefont{Campari}},
  \bibinfo{author}{\bibfnamefont{A.}~\bibnamefont{Belyi}}, \bibnamefont{and}
  \bibinfo{author}{\bibfnamefont{C.}~\bibnamefont{Ratti}},
  \bibinfo{journal}{Physical Review E} \textbf{\bibinfo{volume}{90}},
  \bibinfo{pages}{012811} (\bibinfo{year}{2014}{\natexlab{b}}).

\bibitem[{\citenamefont{Fortunato}(2010)}]{fortunato2010}
\bibinfo{author}{\bibfnamefont{S.}~\bibnamefont{Fortunato}},
  \bibinfo{journal}{Physics Report} \textbf{\bibinfo{volume}{486}},
  \bibinfo{pages}{75} (\bibinfo{year}{2010}).

\bibitem[{\citenamefont{Fortunato and
  Barth{\'e}l{\'e}my}(2007)}]{Fortunato02012007ResolutionLimit}
\bibinfo{author}{\bibfnamefont{S.}~\bibnamefont{Fortunato}} \bibnamefont{and}
  \bibinfo{author}{\bibfnamefont{M.}~\bibnamefont{Barth{\'e}l{\'e}my}},
  \bibinfo{journal}{Proceedings of the National Academy of Sciences}
  \textbf{\bibinfo{volume}{104}}, \bibinfo{pages}{36} (\bibinfo{year}{2007}),
  \eprint{http://www.pnas.org/content/104/1/36.full.pdf+html},
  \urlprefix\url{http://www.pnas.org/content/104/1/36.abstract}.

\bibitem[{\citenamefont{Rosvall and
  Bergstrom}(2007)}]{Rosvall01052007InformationTheoretic}
\bibinfo{author}{\bibfnamefont{M.}~\bibnamefont{Rosvall}} \bibnamefont{and}
  \bibinfo{author}{\bibfnamefont{C.~T.} \bibnamefont{Bergstrom}},
  \bibinfo{journal}{Proceedings of the National Academy of Sciences}
  \textbf{\bibinfo{volume}{104}}, \bibinfo{pages}{7327} (\bibinfo{year}{2007}),
  \eprint{http://www.pnas.org/content/104/18/7327.full.pdf+html},
  \urlprefix\url{http://www.pnas.org/content/104/18/7327.abstract}.

\bibitem[{\citenamefont{Rosvall and Bergstrom}(2008)}]{Infomap}
\bibinfo{author}{\bibfnamefont{M.}~\bibnamefont{Rosvall}} \bibnamefont{and}
  \bibinfo{author}{\bibfnamefont{C.}~\bibnamefont{Bergstrom}},
  \bibinfo{journal}{Proc. Natl. Acad. Sci. USA} \textbf{\bibinfo{volume}{105}},
  \bibinfo{pages}{1118} (\bibinfo{year}{2008}).

\bibitem[{\citenamefont{Karrer and Newman}(2011)}]{Newman2011Stochastic}
\bibinfo{author}{\bibfnamefont{B.}~\bibnamefont{Karrer}} \bibnamefont{and}
  \bibinfo{author}{\bibfnamefont{M.~E.~J.} \bibnamefont{Newman}},
  \bibinfo{journal}{Phys. Rev. E} \textbf{\bibinfo{volume}{83}},
  \bibinfo{pages}{016107} (\bibinfo{year}{2011}),
  \urlprefix\url{http://link.aps.org/doi/10.1103/PhysRevE.83.016107}.

\bibitem[{\citenamefont{Ball et~al.}(2011)\citenamefont{Ball, Karrer, and
  Newman}}]{Newman2011Efficient}
\bibinfo{author}{\bibfnamefont{B.}~\bibnamefont{Ball}},
  \bibinfo{author}{\bibfnamefont{B.}~\bibnamefont{Karrer}}, \bibnamefont{and}
  \bibinfo{author}{\bibfnamefont{M.~E.~J.} \bibnamefont{Newman}},
  \bibinfo{journal}{Phys. Rev. E} \textbf{\bibinfo{volume}{84}},
  \bibinfo{pages}{036103} (\bibinfo{year}{2011}),
  \urlprefix\url{http://link.aps.org/doi/10.1103/PhysRevE.84.036103}.

\bibitem[{\citenamefont{Bickel and Chen}(2009)}]{Bickel2009Nonparametric}
\bibinfo{author}{\bibfnamefont{P.~J.} \bibnamefont{Bickel}} \bibnamefont{and}
  \bibinfo{author}{\bibfnamefont{A.}~\bibnamefont{Chen}},
  \bibinfo{journal}{Proceedings of the National Academy of Sciences}
  \textbf{\bibinfo{volume}{106}}, \bibinfo{pages}{21068}
  (\bibinfo{year}{2009}).

\bibitem[{\citenamefont{Decelle
  et~al.}(2011{\natexlab{a}})\citenamefont{Decelle, Krzakala, Moore, and
  Zdeborov\'a}}]{Decelle2011BlockModel}
\bibinfo{author}{\bibfnamefont{A.}~\bibnamefont{Decelle}},
  \bibinfo{author}{\bibfnamefont{F.}~\bibnamefont{Krzakala}},
  \bibinfo{author}{\bibfnamefont{C.}~\bibnamefont{Moore}}, \bibnamefont{and}
  \bibinfo{author}{\bibfnamefont{L.}~\bibnamefont{Zdeborov\'a}},
  \bibinfo{journal}{Phys. Rev. Lett.} \textbf{\bibinfo{volume}{107}},
  \bibinfo{pages}{065701} (\bibinfo{year}{2011}{\natexlab{a}}),
  \urlprefix\url{http://link.aps.org/doi/10.1103/PhysRevLett.107.065701}.

\bibitem[{\citenamefont{Decelle
  et~al.}(2011{\natexlab{b}})\citenamefont{Decelle, Krzakala, Moore, and
  Zdeborov\'a}}]{Decelle2011BlockModelAsymptotics}
\bibinfo{author}{\bibfnamefont{A.}~\bibnamefont{Decelle}},
  \bibinfo{author}{\bibfnamefont{F.}~\bibnamefont{Krzakala}},
  \bibinfo{author}{\bibfnamefont{C.}~\bibnamefont{Moore}}, \bibnamefont{and}
  \bibinfo{author}{\bibfnamefont{L.}~\bibnamefont{Zdeborov\'a}},
  \bibinfo{journal}{Phys. Rev. E} \textbf{\bibinfo{volume}{84}},
  \bibinfo{pages}{066106} (\bibinfo{year}{2011}{\natexlab{b}}),
  \urlprefix\url{http://link.aps.org/doi/10.1103/PhysRevE.84.066106}.

\bibitem[{\citenamefont{Yan et~al.}(2014)\citenamefont{Yan, Shalizi, Jensen,
  Krzakala, Moore, Zdeborov{\'a}, Zhang, and Zhu}}]{Yan2012ModelSelection}
\bibinfo{author}{\bibfnamefont{X.}~\bibnamefont{Yan}},
  \bibinfo{author}{\bibfnamefont{C.}~\bibnamefont{Shalizi}},
  \bibinfo{author}{\bibfnamefont{J.~E.} \bibnamefont{Jensen}},
  \bibinfo{author}{\bibfnamefont{F.}~\bibnamefont{Krzakala}},
  \bibinfo{author}{\bibfnamefont{C.}~\bibnamefont{Moore}},
  \bibinfo{author}{\bibfnamefont{L.}~\bibnamefont{Zdeborov{\'a}}},
  \bibinfo{author}{\bibfnamefont{P.}~\bibnamefont{Zhang}}, \bibnamefont{and}
  \bibinfo{author}{\bibfnamefont{Y.}~\bibnamefont{Zhu}},
  \bibinfo{journal}{Journal of Statistical Mechanics: Theory and Experiment}
  \textbf{\bibinfo{volume}{2014}}, \bibinfo{pages}{P05007}
  (\bibinfo{year}{2014}).

\bibitem[{\citenamefont{Aldecoa and Mar\`in}(2011)}]{Aldecoa2011Deciphering}
\bibinfo{author}{\bibfnamefont{R.}~\bibnamefont{Aldecoa}} \bibnamefont{and}
  \bibinfo{author}{\bibfnamefont{I.}~\bibnamefont{Mar\`in}},
  \bibinfo{journal}{PLoS ONE} \textbf{\bibinfo{volume}{6}},
  \bibinfo{pages}{e24195} (\bibinfo{year}{2011}),
  \urlprefix\url{http://dx.doi.org/10.1371%2Fjournal.pone.0024195}.

\bibitem[{\citenamefont{Brandes et~al.}(2006)\citenamefont{Brandes, Delling,
  Gaertler, G{\"o}rke, Hoefer, Nikoloski, and Wagner}}]{brandes2006maximizing}
\bibinfo{author}{\bibfnamefont{U.}~\bibnamefont{Brandes}},
  \bibinfo{author}{\bibfnamefont{D.}~\bibnamefont{Delling}},
  \bibinfo{author}{\bibfnamefont{M.}~\bibnamefont{Gaertler}},
  \bibinfo{author}{\bibfnamefont{R.}~\bibnamefont{G{\"o}rke}},
  \bibinfo{author}{\bibfnamefont{M.}~\bibnamefont{Hoefer}},
  \bibinfo{author}{\bibfnamefont{Z.}~\bibnamefont{Nikoloski}},
  \bibnamefont{and} \bibinfo{author}{\bibfnamefont{D.}~\bibnamefont{Wagner}},
  \bibinfo{journal}{arXiv preprint physics/0608255}  (\bibinfo{year}{2006}).
  
\bibitem[{\citenamefont{Dinh et~al.}(2015)\citenamefont{Dinh, Li, and
  Thai}}]{dinh2015network}
\bibinfo{author}{\bibfnamefont{T.~N.} \bibnamefont{Dinh}},
  \bibinfo{author}{\bibfnamefont{X.}~\bibnamefont{Li}}, \bibnamefont{and}
  \bibinfo{author}{\bibfnamefont{M.~T.} \bibnamefont{Thai}}, in
  \emph{\bibinfo{booktitle}{Data Mining (ICDM), 2015 IEEE International
  Conference on}} (\bibinfo{organization}{IEEE}, \bibinfo{year}{2015}), pp.
  \bibinfo{pages}{101--110}.  

\bibitem[{\citenamefont{Cole}(1981)}]{cole1981dominance}
\bibinfo{author}{\bibfnamefont{B.~J.} \bibnamefont{Cole}},
  \bibinfo{journal}{Science} \textbf{\bibinfo{volume}{212}},
  \bibinfo{pages}{83} (\bibinfo{year}{1981}).

\bibitem[{\citenamefont{Connor et~al.}(1992)\citenamefont{Connor, Smolker, and
  Richards}}]{connor1992dolphin}
\bibinfo{author}{\bibfnamefont{R.~C.} \bibnamefont{Connor}},
  \bibinfo{author}{\bibfnamefont{R.~A.} \bibnamefont{Smolker}},
  \bibnamefont{and} \bibinfo{author}{\bibfnamefont{A.~F.}
  \bibnamefont{Richards}}, \bibinfo{journal}{Coalitions and alliances in humans
  and other animals} pp. \bibinfo{pages}{415--443} (\bibinfo{year}{1992}).

\bibitem[{\citenamefont{Clutton-Brock et~al.}(1976)\citenamefont{Clutton-Brock,
  Greenwood, and Powell}}]{clutton1976ranks}
\bibinfo{author}{\bibfnamefont{T.}~\bibnamefont{Clutton-Brock}},
  \bibinfo{author}{\bibfnamefont{P.}~\bibnamefont{Greenwood}},
  \bibnamefont{and} \bibinfo{author}{\bibfnamefont{R.}~\bibnamefont{Powell}},
  \bibinfo{journal}{Ethology} \textbf{\bibinfo{volume}{41}},
  \bibinfo{pages}{202} (\bibinfo{year}{1976}).

\bibitem[{UCI()}]{UCINETIV}
\emph{\bibinfo{title}{Ucinet iv datasets}},
  \urlprefix\url{http://vlado.fmf.uni-lj.si/pub/networks/data/UciNet/UciData.htm}.

\bibitem[{\citenamefont{Lancichinetti et~al.}(2008)\citenamefont{Lancichinetti,
  Fortunato, and Radicchi}}]{LFR}
\bibinfo{author}{\bibfnamefont{A.}~\bibnamefont{Lancichinetti}},
  \bibinfo{author}{\bibfnamefont{S.}~\bibnamefont{Fortunato}},
  \bibnamefont{and} \bibinfo{author}{\bibfnamefont{F.}~\bibnamefont{Radicchi}},
  \bibinfo{journal}{Phys. Rev. E} \textbf{\bibinfo{volume}{78 (4)}},
  \bibinfo{pages}{046110} (\bibinfo{year}{2008}).

\bibitem[{Mor()}]{Moreno}
\emph{\bibinfo{title}{Linton c. freeman's datasets}},
  \urlprefix\url{http://moreno.ss.uci.edu/data.html}.

\bibitem[{\citenamefont{Zachary}(1977)}]{zachary1977ifm}
\bibinfo{author}{\bibfnamefont{W.~W.} \bibnamefont{Zachary}},
  \bibinfo{journal}{Journal of Anthropological Research}
  \textbf{\bibinfo{volume}{33}}, \bibinfo{pages}{452} (\bibinfo{year}{1977}).

\bibitem[{\citenamefont{Lusseau et~al.}(2003)\citenamefont{Lusseau, Schneider,
  Boisseau, Haase, Slooten, and Dawson}}]{Lusseau2003Dolphins}
\bibinfo{author}{\bibfnamefont{D.}~\bibnamefont{Lusseau}},
  \bibinfo{author}{\bibfnamefont{K.}~\bibnamefont{Schneider}},
  \bibinfo{author}{\bibfnamefont{O.~J.} \bibnamefont{Boisseau}},
  \bibinfo{author}{\bibfnamefont{P.}~\bibnamefont{Haase}},
  \bibinfo{author}{\bibfnamefont{E.}~\bibnamefont{Slooten}}, \bibnamefont{and}
  \bibinfo{author}{\bibfnamefont{S.~M.} \bibnamefont{Dawson}},
  \bibinfo{journal}{Behavioral Ecology and Sociobiology}
  \textbf{\bibinfo{volume}{54}}, \bibinfo{pages}{396} (\bibinfo{year}{2003}),
  \urlprefix\url{http://dx.doi.org/10.1007/s00265-003-0651-y}.

\bibitem[{\citenamefont{Knuth}(1993)}]{Knuth1993GraphBase}
\bibinfo{author}{\bibfnamefont{D.~E.} \bibnamefont{Knuth}},
  \emph{\bibinfo{title}{{The Stanford GraphBase: a platform for combinatorial
  computing}}} (\bibinfo{publisher}{Addison-Wesley}, \bibinfo{year}{1993}),
  \urlprefix\url{http://www-cs-staff.stanford.edu/\~{}uno/sgb.html}.

\bibitem[{\citenamefont{White et~al.}(1986)\citenamefont{White, Southgate,
  Thomson, and Brenner}}]{White12111986}
\bibinfo{author}{\bibfnamefont{J.~G.} \bibnamefont{White}},
  \bibinfo{author}{\bibfnamefont{E.}~\bibnamefont{Southgate}},
  \bibinfo{author}{\bibfnamefont{J.~N.} \bibnamefont{Thomson}},
  \bibnamefont{and} \bibinfo{author}{\bibfnamefont{S.}~\bibnamefont{Brenner}},
  \bibinfo{journal}{Philosophical Transactions of the Royal Society of London.
  B, Biological Sciences} \textbf{\bibinfo{volume}{314}}, \bibinfo{pages}{1}
  (\bibinfo{year}{1986}),
  \eprint{http://rstb.royalsocietypublishing.org/content/314/1165/1.full.pdf+html},
  \urlprefix\url{http://rstb.royalsocietypublishing.org/content/314/1165/1.abstract}.

\bibitem[{\citenamefont{Isella et~al.}(2011)\citenamefont{Isella, Stehl{\'e},
  Barrat, Cattuto, Pinton, and Van~den Broeck}}]{isella2011s}
\bibinfo{author}{\bibfnamefont{L.}~\bibnamefont{Isella}},
  \bibinfo{author}{\bibfnamefont{J.}~\bibnamefont{Stehl{\'e}}},
  \bibinfo{author}{\bibfnamefont{A.}~\bibnamefont{Barrat}},
  \bibinfo{author}{\bibfnamefont{C.}~\bibnamefont{Cattuto}},
  \bibinfo{author}{\bibfnamefont{J.-F.} \bibnamefont{Pinton}},
  \bibnamefont{and} \bibinfo{author}{\bibfnamefont{W.}~\bibnamefont{Van~den
  Broeck}}, \bibinfo{journal}{Journal of theoretical biology}
  \textbf{\bibinfo{volume}{271}}, \bibinfo{pages}{166} (\bibinfo{year}{2011}).

\bibitem[{\citenamefont{Gleiser and Danon}(2003)}]{Gleiser2003Jazz}
\bibinfo{author}{\bibfnamefont{P.~M.} \bibnamefont{Gleiser}} \bibnamefont{and}
  \bibinfo{author}{\bibfnamefont{L.}~\bibnamefont{Danon}},
  \bibinfo{journal}{Advances in Complex Systems} \textbf{\bibinfo{volume}{06}},
  \bibinfo{pages}{565} (\bibinfo{year}{2003}),
  \eprint{http://www.worldscientific.com/doi/pdf/10.1142/S0219525903001067},
  \urlprefix\url{http://www.worldscientific.com/doi/abs/10.1142/S0219525903001067}.

\end{thebibliography}
\end{document}